\documentclass[aps,prl,reprint,showpacs,footinbib,amsmath,amssymb,amsfonts,floatfix,superscriptaddress]{revtex4-1}

\usepackage{graphicx}
\usepackage{epsfig}
\usepackage{color}

\begin{document}


\title{Recycling controls membrane domains}



\author{S. Alex Rautu}
\affiliation{Department of Physics, University of Warwick, Coventry, CV4 7AL, United Kingdom}

\author{George Rowlands}
\affiliation{Department of Physics, University of Warwick, Coventry, CV4 7AL, United Kingdom}

\author{Matthew S. Turner}
\affiliation{Department of Physics, University of Warwick, Coventry, CV4 7AL, United Kingdom}
\affiliation{Centre for Complexity Science, University of Warwick, Coventry CV4 7AL, United Kingdom}


\date{\today}

\begin{abstract}

We study the coarsening of strongly microphase separated membrane domains in the presence of recycling of material. We study the dynamics of the domain size distribution under both scale-free and size-dependent recycling. Closed form solutions to the steady state distributions and its associated central moments are obtained in both cases. Moreover, for the size-independent case, the~time evolution of the moments is analytically calculated, which provide us with exact results for their corresponding relaxation times. Since these moments and relaxation times are measurable quantities, the biophysically significant free parameters in our model may be determined by comparison with experimental data.

\end{abstract}



\maketitle


Biomembranes are highly dynamic two-dimensional systems, consisting of many different lipids and proteins \cite{Engelman2005}, which are continuously exchanged with the rest of the living cell by the secretion and absorption of vesicles of approximately $50$--$200$~nm in diameter~\cite{Kobayashi1998}. This recycling of cell membranes leads to a complete turnover of its constituents in about 12 minutes \cite{Hao2000}. In addition, the membrane components are found to be inhomogeneously distributed \cite{Arumugam2015}, where certain lipids and proteins cluster into small-scale domains with a diameter of few tens of nanometers, which are sometimes referred to as lipid rafts~\cite{Simons1997, Simons2000, Simons2011}. Such supermolecular structures are free to diffuse throughout the membrane, coalescing into larger domains as they meet~\cite{Lavi2007}. They are  believed to be involved in controlling many biological processes, such as signal transduction, protein sorting, and endocytosis \cite{Simons2011}. Although there is a growing evidence of their existence and biological importance to living cells~\cite{Destainville2008, Ying2009, Robson2012, Brown2006, Mouritsen2011}, there are still many unanswered questions concerning both their origin and nature \cite{Komura2014, Pike2006, Leslie2011, Munro2003, Veatch2005, Pike2009}. As lipid phase separation can occur in model systems of multi-component membranes, lipid rafts have often been linked to micro-phase separated lipid domains \cite{Veatch2005}. Nonetheless, the size of these domains are much larger than those observed in cells.  This is expected as phase separation in a two-component mixture manifests itself by the appearance of separated domains (below a critical temperature), which then grow until they reach the size of the system, without any intermediate stable sizes~\cite{Bray1994}.  This argument makes the existence of lipid rafts somewhat surprising from a physical point of view, as they posses a characteristic size that is much smaller than the typical diameter of cells. 

\begin{figure}[b!]\includegraphics[width=\columnwidth]{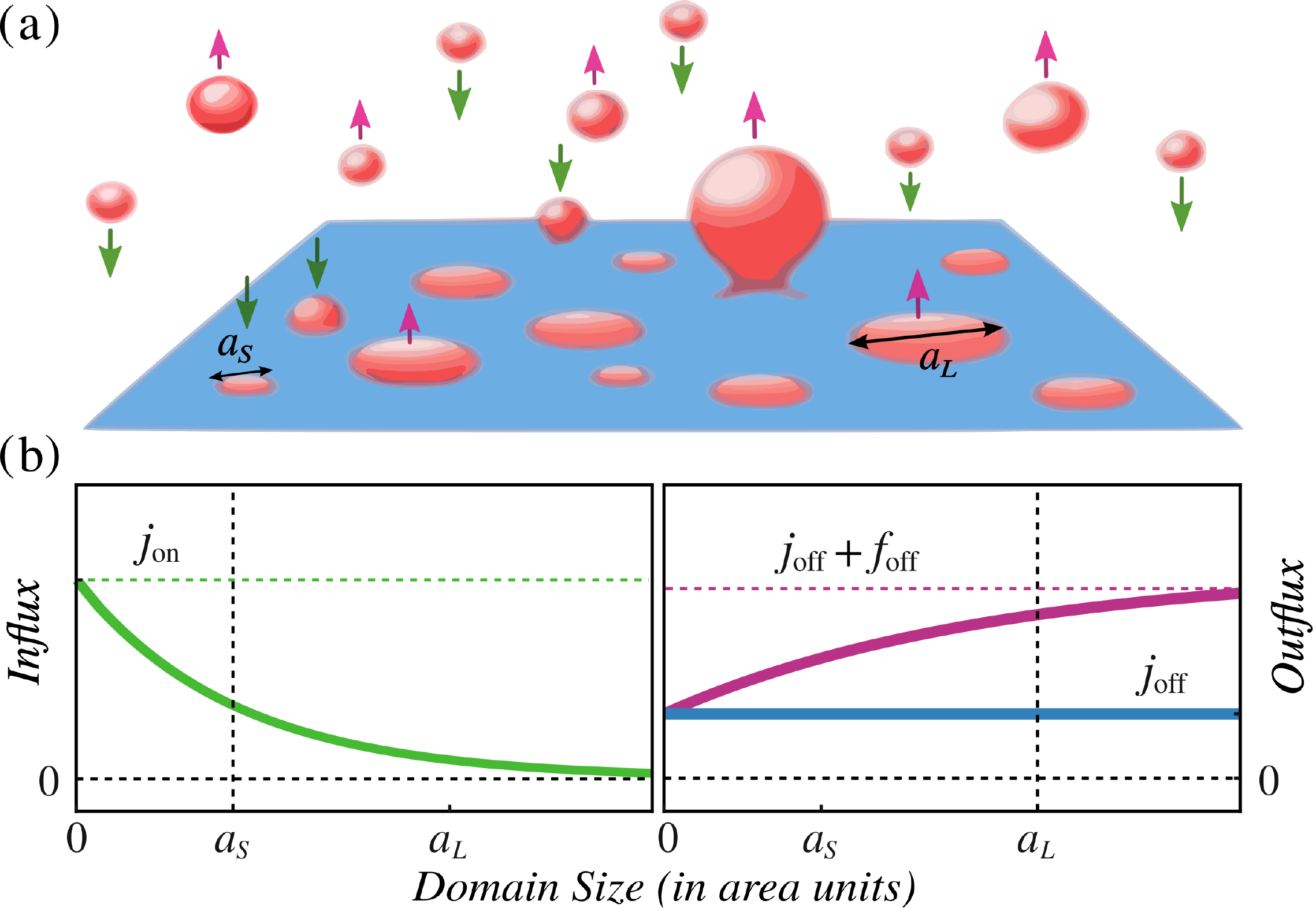}
\caption{\label{fig:1} (color on-line) (a) Schematic diagram of a planar membrane that is composed of two lipid species, depicted by blue and red colors. The latter constituent phase-separates into membrane domains of different sizes that range between the characteristic areas $a_{\scriptscriptstyle S}\simeq10\textnormal{ nm}^2$ and $a_{\scriptscriptstyle L}\simeq10^4\textnormal{ nm}^2$. In addition to their in-plane diffusive dynamics, the membrane is subjected to a continuous recycling, where the lipid domains are constantly brought to (green downward arrows) and removed from (pink upward arrows) the membrane through the transport of various endosomes or vesicles. (b) Schematic of the membrane recycling, where single domains are injected into the membrane at random with a flux $j_{\scriptscriptstyle \textnormal{on}}\exp\!\left(-a/a_{\scriptscriptstyle S}\right)/\,a_{\scriptscriptstyle S}$. At the same time, domains are randomly removed from the membrane with a constant rate $j_{\scriptscriptstyle \textnormal{off}}$ as well as an explicit size-dependent removal rate, $f_{\scriptscriptstyle \textnormal{off}}\left[1-\exp\left(-a/a_{\scriptscriptstyle L}\right)\right]$, which shows a linear regime up to $a_{\scriptscriptstyle L}$, plateauing to $f_{\scriptscriptstyle \textnormal{off}}+j_{\scriptscriptstyle \textnormal{off}}$ for large $a$.  }\end{figure}

In this Letter, we study the formation and regulation of stable nano-scale domains by membrane recycling. We develop a continuum theory of the domain dynamics under a continuous exchange of membrane components with an external (or internal) reservoir \cite{Turner2005, Foret2005}. We consider an infinite planar membrane populated by two lipid species, where the lipids undergo phase separation, giving rise to domains of various sizes, which are surrounded by the other membrane component. Herein, the domain scission events are assumed to be rare, corresponding to a regime of large line tension \cite{Turner2005}. The latter characterizes the energy cost for having a finite interface between the different phases. For phase separating lipids the regime of most interest is high line tension, while the low tension case resembles a gas of non-interacting (mostly monomeric) domains. Consequently, the kinetics of the membrane domains is mostly dominated by the fusion events \cite{Turner2005}. Thus, the mean-field dynamics of the distribution of domain sizes under continuous recycling is governed by the following master equation \cite{Leyvraz2003}:
\begin{align}
  \label{eqn:master-equation}
  \frac{\mathrm{d}\mathcal{P}}{\mathrm{d}t} &= 
  \mathcal{R}(a) -\int^{\infty}_{0}\mathcal{G}\left(a,\,a'\right)\mathcal{P}\left(a,\,t\right)\mathcal{P}\left(a'\!,\,t\right)\,\mathrm{d}a' \notag\\[-1pt]
  &\,+ \frac{1}{2}\,\int_{0}^{\,a}\mathcal{G}\left(a,\,a'\right)\mathcal{P}\left(a'\!,\,t\right)\,\mathcal{P}\left(a-a'\!,\,t\right)\,\mathrm{d}a',
\end{align} where $\mathcal{P}(a,t)$ represents a density function at time $t$ for the number-per-area of domains of size $a$ (in area units). Hereinafter, we assume that two lipid domains coalesce whenever they come into contact through diffusion, so that the kernel $\mathcal{G}(a,a')$ in Eq.~(\ref{eqn:master-equation}) can be regarded as a constant proportional to the diffusion coefficient $D$ of lipid rafts \footnote{Due to the logarithmic dependence of the diffusion coefficient on the size $a$ of the lipid domains (according to Saffman-Delbruck theory \cite{Saffman1975}), namely $D\sim\log(1/a)$, we neglect the size-dependence of the fusion rate as result.}. Since this is the only parameter that describes the intramembrane dynamics, $\mathcal{G}$ is chosen to be identically~$D\simeq10^5\textnormal{ nm}^2/\,\textnormal{s}$, which fixes the time scale in our model. Lastly, $\mathcal{R}(a)$ is a function \footnote{For the sake of brevity, we omit the explicit time dependence of this function, arising from the density~$\mathcal{P}(a,t)$.} that controls the lipid recycling (as shown in Fig.~\ref{fig:1}), namely
\begin{equation}
\label{eqn:scheme}
\mathcal{R}(a)=j_{\scriptscriptstyle \textnormal{on}}\,\frac{e^{-a/a_{\scriptscriptstyle S}}}{a_{\scriptscriptstyle S}} - \left[j_{\scriptscriptstyle \textnormal{off}} + f_{\scriptscriptstyle \textnormal{off}}\,\big{(}1-e^{-a/a_{\scriptscriptstyle L}}\big{)}\!\right]\!\mathcal{P}(a),
\end{equation} where single domains are brought to the membrane at random with a flux $j_{\scriptscriptstyle \textnormal{on}}$ and with a size drawn from an normalized exponential distribution. Here, $a_{\scriptscriptstyle S}\simeq10\textnormal{ nm}^2$ is the characteristic size of domains that are injected into the membrane. Moreover, entire domains are stochastically removed irrespective of their size with a constant rate $j_{\scriptscriptstyle \textnormal{off}}$. In addition to this, an explicit size-dependent outward flux is included, where the removal rate $f_{\scriptscriptstyle \textnormal{off}}$ is exponentially small for domains sizes $a\lesssim a_{\scriptscriptstyle L} \simeq10^4\textnormal{ nm}^2$. This recycles mainly at large or small scales, depending on the sign choice of $f_{\scriptscriptstyle \textnormal{off}}$. The former scenario is perhaps of greater biological relevance, due to the size associated with the endosomes, which are vapoured in the phase separated component~\cite{Simons2011}. This has also been observed for the recycling of E-cadherin {\it in vivo}, where dynamin-dependent endocytosis targets large domains, inducing a sharp cutoff past a critical size~\cite{TruongQuang2013}.

\begin{figure}[t]\includegraphics[width=\columnwidth]{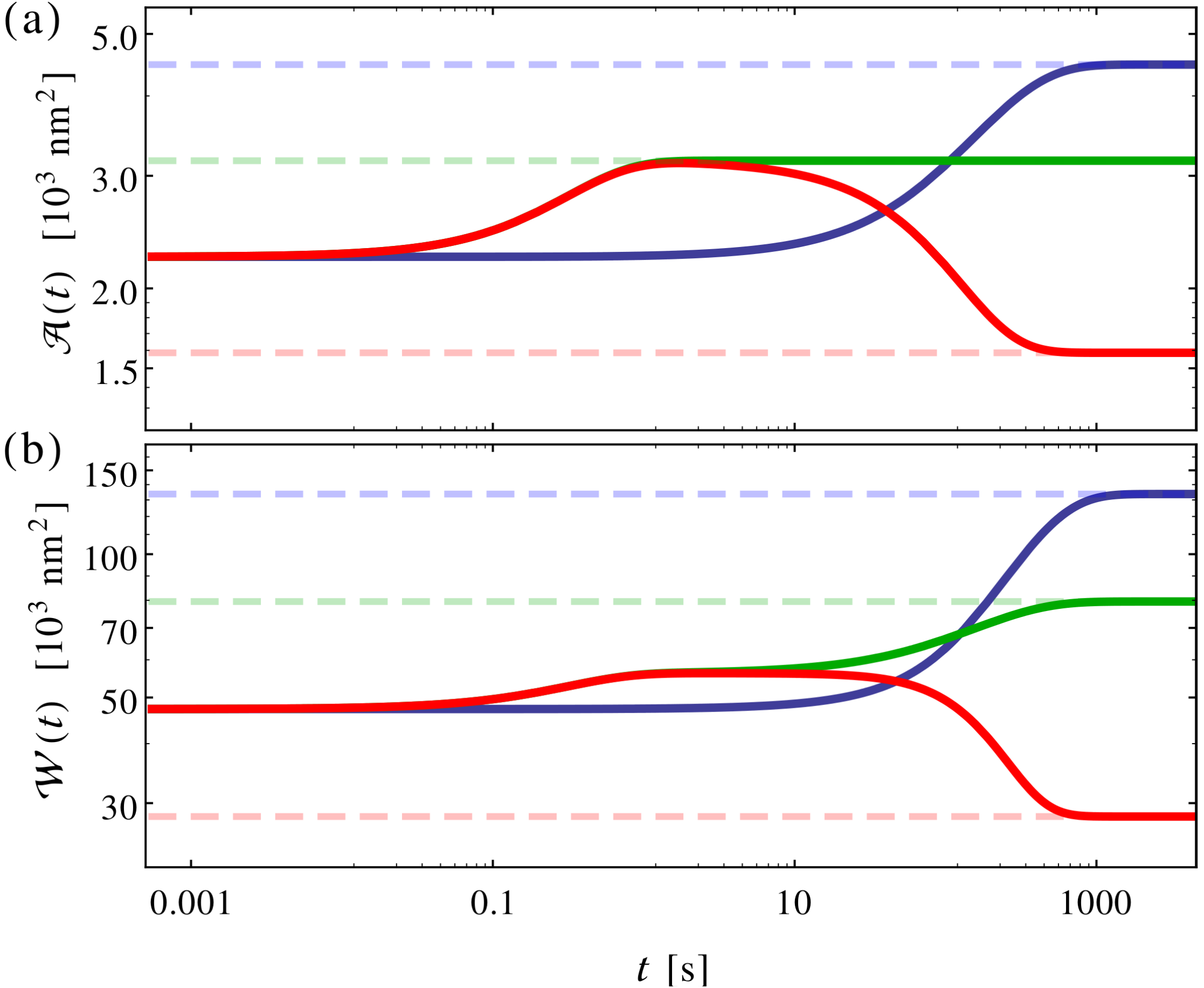}
\caption{
\label{fig:2} 
(color on-line) (a) Time evolution of the mean domain size~$\mathcal{A}(t)$, and (b) its associated standard deviation~$\mathcal{W}(t)$. The~initial boundary conditions are chosen by considering a scenario where a step-change at $t=0$ is made in either $J_{\scriptscriptstyle \textnormal{on}}$ or $J_{\scriptscriptstyle \textnormal{off}}$ after the system has reached its steady-state configuration. Choosing the typical physiological values, $J_{\scriptscriptstyle \textnormal{on}}=10^{-7}$ and $J_{\scriptscriptstyle \textnormal{off}}=10^{-6}$, we make the following step-like changes in the recycling rates: 50\% decrease in $J_{\scriptscriptstyle \textnormal{off}}$ (blue); 50\% decrease in $J_{\scriptscriptstyle \textnormal{on}}$ (red); and 50\% decrease in both $J_{\scriptscriptstyle \textnormal{on}}$ and $J_{\scriptscriptstyle \textnormal{off}}$ (green). Due to these perturbations the system reaches a new steady-state after some transient time, as shown by the dashed lines (same color convention). 
}\end{figure}

To understand the solution of Eq.~(\ref{eqn:master-equation}) we first focus on the size-independent recycling scheme, namely $f_{\scriptscriptstyle \textnormal{off}}=0$. This proves to be tractable in Laplace space~\cite{DLMF}, where we define the dimensionless (integral transform) function $\hat{\mathcal{P}}\!\left(\lambda,t\right) = a_{\scriptscriptstyle S}\int^{\infty}_{0}\mathcal{P}\!\left(a,t\right) e^{-a\hspace{1pt}\lambda/a_{\scriptscriptstyle S}}\,\mathrm{d}a$. For clarity it is helpful to define the following rescaled quantities at the outset: $\tau = t\,D/a_{\scriptscriptstyle S}$, $J_{\scriptscriptstyle \textnormal{on}} = j_{\scriptscriptstyle \textnormal{on}}\,a^2_{\scriptscriptstyle S}/D$, and $J_{\scriptscriptstyle \textnormal{off}} = j_{\scriptscriptstyle \textnormal{off}}\,a_{\scriptscriptstyle S}/D$. This leads a nonlinear differential equation of the form:
\begin{equation}
  \label{eqn:diff-generating-fct}
  \frac{\mathrm{d}\hat{\mathcal{P}}}{\mathrm{d}\tau} = \frac{J_{\scriptscriptstyle \textnormal{on}}}{1 + \lambda} - \left[\rho\left(\tau\right)+J_{\scriptscriptstyle \textnormal{off}}\right]\hat{\mathcal{P}}\!\left(\lambda,\tau\right) + \frac{1}{2}\,\hat{\mathcal{P}}^2\!\left(\lambda,\tau\right),
\end{equation} where $\rho(\tau) = \hat{\mathcal{P}}\!\left(\lambda=0,\,\tau\right)$ is the total number-per-area of domains (non-dimensionalized by $a_{\scriptscriptstyle S}$). By evaluating Eq.~(\ref{eqn:diff-generating-fct}) at $\lambda=0$, we have that
\begin{equation}
  \label{eqn:diff-rho}
  \frac{\mathrm{d}\rho}{\mathrm{d}\tau} = J_{\scriptscriptstyle \textnormal{on}} - J_{\scriptscriptstyle \textnormal{off}}\,\rho(\tau) - \frac{1}{2}\,\rho^2(\tau),
\end{equation} which can be solved by separation of variables. By using the initial condition $\rho_0 = \rho(\tau=0)$, the solution to Eq.~(\ref{eqn:diff-rho})  can be written as
\begin{equation}
  \label{eqn:rho-sol}
  \rho(\tau) = \mathcal{Q}_{\scriptscriptstyle \infty}\,\frac{\left(\rho_0+J_{\scriptscriptstyle \textnormal{off}}\right)+\mathcal{Q}_{\scriptscriptstyle \infty}\tanh\!\left[\frac{\tau \mathcal{Q}_{\scriptscriptstyle \infty}}{2}\right]}{ \left(\rho_0+J_{\scriptscriptstyle \textnormal{off}}\right)\tanh\!\left[\frac{\tau \mathcal{Q}_{\scriptscriptstyle \infty}}{2}\right]+\mathcal{Q}_{\scriptscriptstyle \infty}}-J_{\scriptscriptstyle \textnormal{off}},
\end{equation}  where $\mathcal{Q}_{\scriptscriptstyle \infty} =\sqrt{J^{\hspace{0.5pt}2}_{\scriptscriptstyle \textnormal{off}} + 2J_{\scriptscriptstyle \textnormal{on}}}$ \footnote{See Supplemental Material, which includes Refs. \cite{Leyvraz2003, Turner2005, DLMF, Abramowitz1965}, 
for further supporting theoretical calculations.}. In order to find $\hat{\mathcal{P}}(\lambda,\tau)$, we define a new function $\psi(\lambda,\tau) = \rho(\tau) - \hat{\mathcal{P}}(\lambda,\tau)$, which by substitution into Eq.~(\ref{eqn:diff-generating-fct}) yields
\begin{equation}
  \label{eqn:diff-psi}
  \frac{\mathrm{d}\psi}{\mathrm{d}\tau} = \frac{\lambda\,J_{\scriptscriptstyle \textnormal{on}}}{1+\lambda} - J_{\scriptscriptstyle \textnormal{off}}\,\psi(\lambda,\tau) -\frac{1}{2}\,\psi^2(\lambda,\tau).
\end{equation} This has the same form as Eq.~(\ref{eqn:diff-rho}), and its solution is
\begin{equation}
  \label{eqn:psi-sol}
  \psi(\lambda,\tau) = \mathcal{Q}_\lambda\,\frac{\left(\psi_0(\lambda)+J_{\scriptscriptstyle \textnormal{off}}\right)+\mathcal{Q}_\lambda\tanh\!\left[\frac{\tau \mathcal{Q}_\lambda}{2}\right]}{\left(\psi_0(\lambda)+J_{\scriptscriptstyle \textnormal{off}}\right)\tanh\!\left[\frac{\tau \mathcal{Q}_\lambda}{2}\right]+\mathcal{Q}_\lambda}-J_{\scriptscriptstyle \textnormal{off}},
\end{equation} where $\psi_0(\lambda) = \psi(\lambda,\,\tau=0)$ and $\mathcal{Q}_\lambda= \sqrt{J^{\hspace{0.5pt}2}_{\scriptscriptstyle \textnormal{off}} + \frac{2\lambda}{1+\lambda}\,J_{\scriptscriptstyle \textnormal{on}}}$. Higher order moments of $\mathcal{P}$ can be determined by differentiating $\hat{\mathcal{P}}(\lambda,\tau)$, or equivalently $-\psi(\lambda,\tau)$, with respect to $\lambda$ and then evaluating at $\lambda=0$. Particularly, its first moment $ \phi(\tau) = \!\int^{\infty}_{0}\! a\,\mathcal{P}(a,\tau)\,\mathrm{d}a = -\frac{\mathrm{d}}{\mathrm{d}\lambda}\hat{\mathcal{P}}\!\left(\lambda=0,\,\tau\right)$, which corresponds to the area-fraction of domains, is found to be
\begin{equation}
   \label{eqn:1stmoment}
    \phi(\tau) = \frac{J_{\text{on}}}{J_{\text{off}}}\left[1-e^{- \tau J_{\scriptscriptstyle \textnormal{off}}} \left(1-\frac{J_{\scriptscriptstyle \textnormal{off}}}{J_{\text{on}}}\,\phi_0 \right)\right]\!,
\end{equation} where $\phi_0 = \frac{\mathrm{d}}{\mathrm{d}\lambda}\psi_0(\lambda=0)$. This allows us to find the time evolution of the mean domain size $\mathcal{A}(\tau) = a_{\scriptscriptstyle S}\,\phi(\tau)/\rho(\tau)$, as shown in Fig.~\ref{fig:2}\,(a). Similarly, the second moment, i.e.\ $\sigma(\tau) = \frac{\mathrm{d}^2}{\mathrm{d}\lambda^2}\hat{\mathcal{P}}\!\left(\lambda=0,\,\tau\right)$, can be exactly computed \cite{Note3}, which together with lower raw-moments it gives us the full dynamics of the standard deviation $\mathcal{W}(\tau)$ associated to the domain size distribution $\mathcal{P}$, see Fig.~\ref{fig:2}\,(b). 

The solutions in  Fig.~\ref{fig:2} show how the entire system will relax after an initial perturbation in the recycling rates. Such an assay could plausibility be performed experimentally by up-regulating or knocking down key elements of the synthesis or endocytic pathway. Since the relaxation times of the central moments can be on the order of tens of minutes (cf.\ Fig.~\ref{fig:2}), they can be experimentally measurable. This allows us to estimate $J_{\text{on}}$ and $J_{\text{off}}$ by comparison with the decay rates of (\ref{eqn:rho-sol}) and (\ref{eqn:1stmoment})
. Another method that can be used to find the recycling rates is to measure the steady-state values of both $\rho(\tau)$ and $\phi(\tau)$. These are given by their limits as time $\tau\to\infty$, namely $\rho_{\scriptscriptstyle \infty} = \mathcal{Q}_{\scriptscriptstyle \infty} - J_{\scriptscriptstyle \textnormal{off}}$ and $\phi_{\scriptscriptstyle \infty} = J_{\scriptscriptstyle \textnormal{on}}/J_{\scriptscriptstyle \textnormal{off}}$, respectively. These methods illustrate the predictive power of our model.

\begin{figure}[t]\includegraphics[width=\columnwidth]{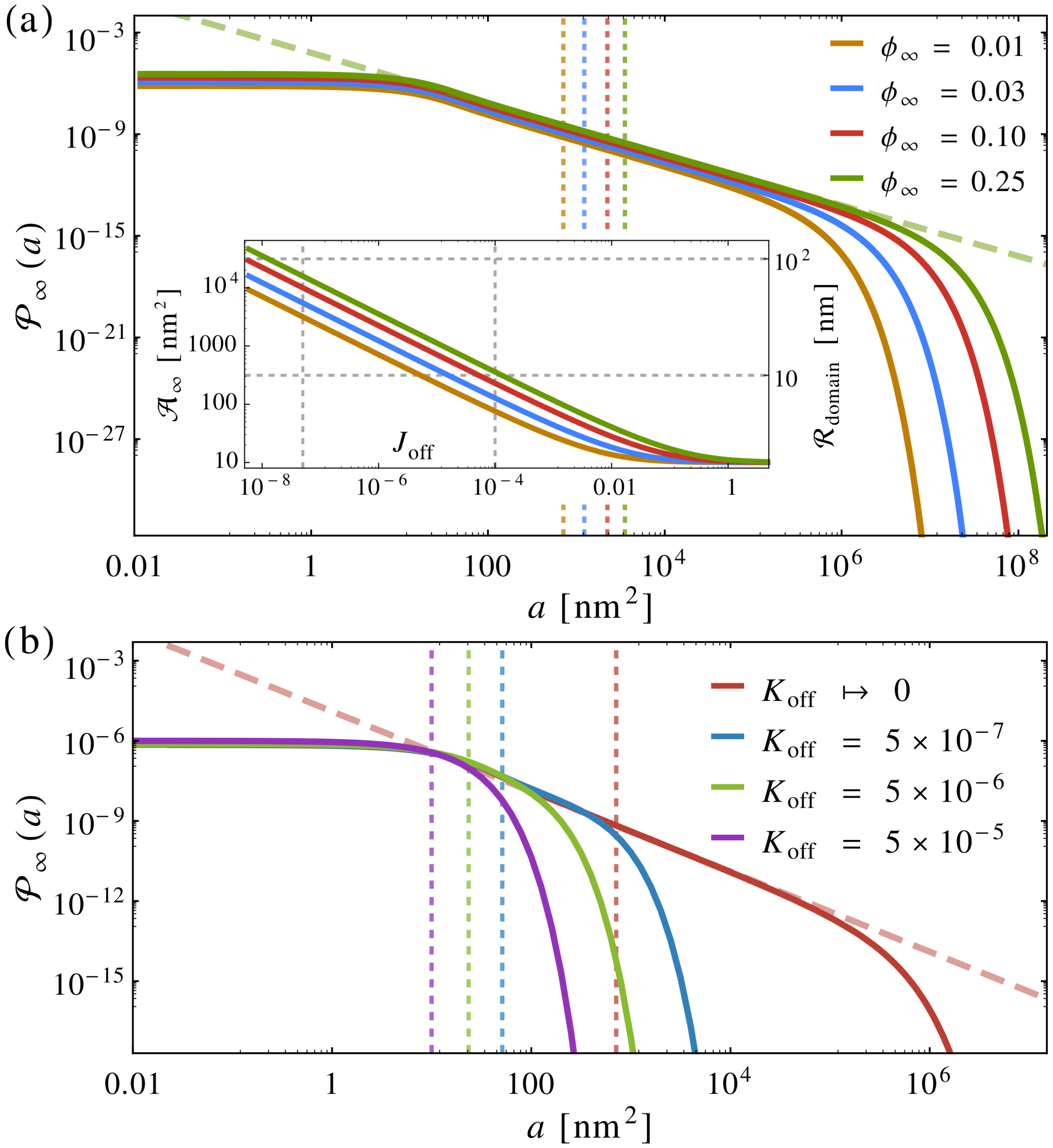}
\caption{\label{fig:3} 
(color on-line) (a) Log-log plot of the steady-state size distribution~$\mathcal{P}_{\scriptscriptstyle \infty}(a)$ for the size-independent recycling at a fixed area coverage $\phi_{\scriptscriptstyle \infty}=J_{\scriptscriptstyle \textnormal{on}}/J_{\scriptscriptstyle \textnormal{off}}$. The outward recycling rate $J_{\scriptscriptstyle \textnormal{off}}$ is chosen such that the average area $\mathcal{A}_{\scriptscriptstyle \infty}$ from the distribution $\mathcal{P}_{\scriptscriptstyle \infty}(a)$ (as indicated by the vertical dashed lines) corresponds to the typical size of phase-separated domains observed in living cells; namely, $J_{\scriptscriptstyle \textnormal{off}} = 10^{-6}$. The latter is found by plotting the mean area $\mathcal{A}_{\scriptscriptstyle \infty}$ as a function of $J_{\scriptscriptstyle \textnormal{off}}$ at a constant area-fraction~$\phi_{\scriptscriptstyle \infty}$ as shown in the inset plot (same color convention). The gray dashed lines represent the upper and lower bounds to the physiological values of $J_{\scriptscriptstyle \textnormal{off}}$ and the membrane area~$\pi\,\mathcal{R}^2_\textnormal{domain}$ of the nano-scale domains (or raft-like structures). The distribution~$\mathcal{P}_{\scriptscriptstyle \infty}(a)$ displays an exponential cut-off for large $a$, and a power-law behavior, $\mathcal{P}_{\scriptscriptstyle \infty}(a)\sim a^{-3/2}$, for intermediate values ($a\!\gtrsim\!a_{\scriptscriptstyle S}$), as displayed by the green dashed line where $\phi_{\scriptscriptstyle \infty}=0.25$. (b) The steady-state distribution for a size-dependent recycling scheme, parametrized by the non-dimensional rate~$K_{\scriptscriptstyle \textnormal{off}}$ (see text), which retrieves the size-independent case in the limit $K_{\scriptscriptstyle \textnormal{off}}=0$. Herein, $J_{\scriptscriptstyle \textnormal{on}}=10^{-7}$ and $J_{\scriptscriptstyle \textnormal{off}}=10^{-6}$. This shows that even a small value of $K_{\scriptscriptstyle \textnormal{off}}$ decreases the size at which the domains are exponentially recycled, reducing the power-law regime (see red dashed line). For larger values of $K_{\scriptscriptstyle \textnormal{off}}\gg J_{\scriptscriptstyle \textnormal{off}}$, the steady-state distribution tends to a purely exponential function. The vertical dashed lines represent the mean domain sizes (same colors).
}\end{figure}

Moreover, the steady-state value of $\hat{\mathcal{P}}(\lambda,\tau)$ is found to be $\hat{\mathcal{P}}_{\scriptscriptstyle \infty}(\lambda) = \mathcal{Q}_{\scriptscriptstyle \infty} - \mathcal{Q}_\lambda$. This can be inverse Laplace transformed, and a closed-form solution to the steady-state distribution can be derived as follows~\cite{Note3}:
\begin{equation}
 \label{eqn:sol-steady-dist}
 \mathcal{P}_{\scriptscriptstyle \infty}(a) = \frac{J_{\scriptscriptstyle \textnormal{on}}\,e^{-a\left(1-\Omega\right)/a_{\scriptscriptstyle S}}}{a^2_{\scriptscriptstyle S}\,\mathcal{Q}_{\scriptscriptstyle \infty}}\left[I_0\!\left(\frac{a\,\Omega}{a_{\scriptscriptstyle S}}\right) - I_1\!\left(\frac{a\,\Omega}{a_{\scriptscriptstyle S}}\right)\right]\!,
\end{equation} where $\Omega = J_{\scriptscriptstyle \textnormal{on}}/\mathcal{Q}^2_{\scriptscriptstyle \infty}$, and also $I_1$ and $I_0$ are the modified Bessel functions of the first kind of order one and zero, respectively~\cite{Abramowitz1965}.  Fig.~\ref{fig:3}\,(a) shows a few plots of Eq.~(\ref{eqn:sol-steady-dist}) for physiologically reasonable values of $J_{\text{on}}$ and $J_{\text{off}}$. This further illustrates that small finite size domains can be obtained within this model for a wide range of recycling rates (see inset plot of Fig.~\ref{fig:3}). $\mathcal{P}_{\scriptscriptstyle \infty}(a)$ shows a power-law behavior with a exponential cut-off \cite{Turner2005, Connaughton2010}, which is also asserted by the asymptotic expansion of Eq.~(\ref{eqn:sol-steady-dist}), namely 
\begin{equation}
\label{eqn:power-law}
\mathcal{P}_{\scriptscriptstyle \infty}(a)\simeq \frac{e^{-a/a_{\scriptscriptstyle C}}}{a^{3/2}}\,\sqrt{\frac{a_{\scriptscriptstyle S} J_{\scriptscriptstyle \textnormal{on}}}{8\pi\Omega^2}},   
\end{equation} where $a_{\scriptscriptstyle C} = a_{\scriptscriptstyle S}\left(1+2\,J_{\scriptscriptstyle \textnormal{on}}/ J^{\hspace{0.5pt}2}_{\scriptscriptstyle \textnormal{off}}\right)\simeq 4\,\mathcal{A}^2_{\scriptscriptstyle \infty}/a_{\scriptscriptstyle S}$, with $\mathcal{A}_{\scriptscriptstyle \infty}$ as the steady-state value of the average domain size. Since the critical size $a_{\scriptscriptstyle C}\gg\mathcal{A}_{\scriptscriptstyle \infty}$, the mean size will mostly lie within the power-law regime. This scaling has been observed in the cluster size distribution of E-cadherin~\cite{TruongQuang2013}.

\begin{figure*}[t]\includegraphics[width=\textwidth]{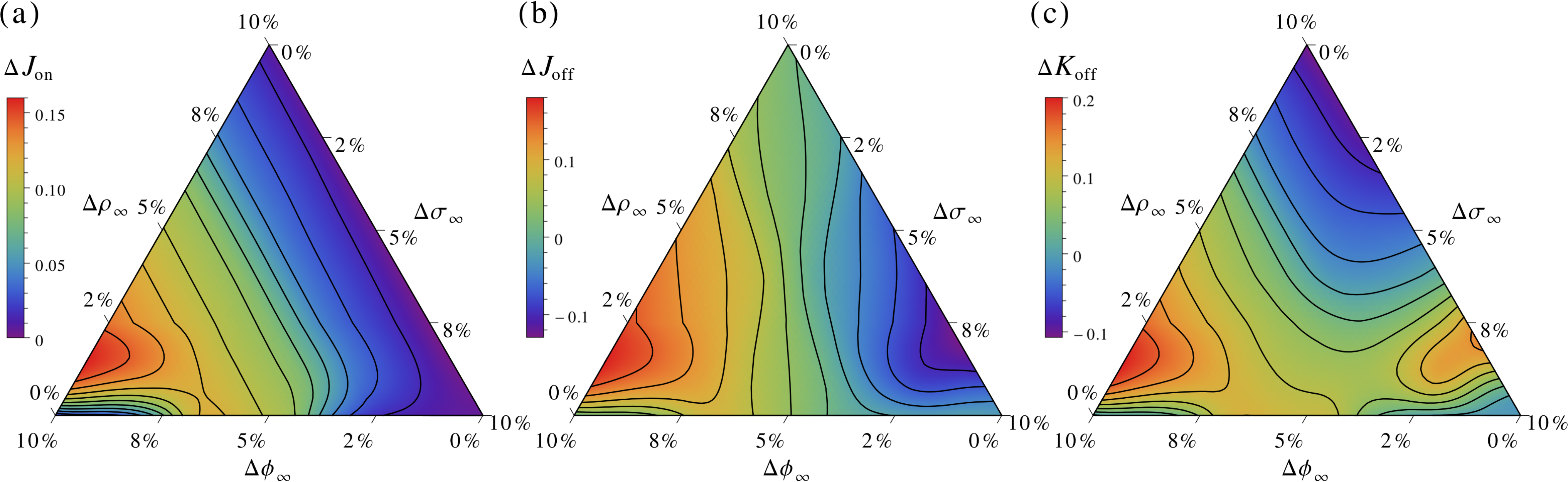}
\caption{\label{fig:4} (color on-line) Ternary plots of the fractional changes (indicated by the prefix $\Delta$) in (a) the injection rate $J_{\scriptscriptstyle \textnormal{on}} = 10^{-5}$, (b) the size-independent removal rate $J_{\scriptscriptstyle \textnormal{off}} = 10^{-4}$, and (c) the size-dependent outward rate $K_{\scriptscriptstyle \textnormal{off}} = 10^{-8}$, which result from the corresponding fractional changes (or errors) in the statistical moments of the steady-state domain size distribution; namely, the total number-per-area~$\rho_{\scriptscriptstyle \infty}$ of domains, their area-fraction~$\phi_{\scriptscriptstyle \infty}$, and the second moment~$\sigma_{\scriptscriptstyle \infty}$ of the distribution, which are computed in the linearized regime (see main text) for the values of the recycling rates mentioned above.}\end{figure*}

We now consider the size-dependent recycling scheme, where $f_{\scriptscriptstyle \textnormal{off}}\neq0$ in Eq.~(\ref{eqn:scheme}). By Laplace transforming the master equation and non-dimensionalizing as done before, we derive a similar expression to Eq.~(\ref{eqn:diff-generating-fct}) with an additional term on the right-hand-side of the equation that is given by $\mathcal{F}(\lambda,\tau) = F_{\scriptscriptstyle \textnormal{off}}\big[\hat{\mathcal{P}}\!\left(\lambda_\star + \lambda,\,\tau\right)-\hat{\mathcal{P}}\!\left(\lambda,\,\tau\right)\!\big]\!$, with $F_{\scriptscriptstyle \textnormal{off}} = f_{\scriptscriptstyle \textnormal{off}}\,a_{\scriptscriptstyle S}/D$ and $\lambda_\star = a_{\scriptscriptstyle S}/a_{\scriptscriptstyle L}$. Thus, this leads to a nonlinear differential equation with a recurrence-like relation for the continuous variable $\lambda$. Such equations are difficult to solve exactly or even numerically. However, further analytical progress can be made by assuming that the ratio $\lambda_\star\ll1$ (typically $\lambda_\star\simeq10^{-3}$), which yields that $\mathcal{F}(\lambda,\tau)= K_{\scriptscriptstyle \textnormal{off}}\,\frac{\mathrm{d}}{\mathrm{d}\lambda}\hat{\mathcal{P}}\!\left(\lambda,\tau\right)$, with $K_{\scriptscriptstyle \textnormal{off}} = \lambda_\star\,F_{\scriptscriptstyle \textnormal{off}}$. Hence, the steady-state equation associated to $\hat{\mathcal{P}}(\lambda,\tau)$ is given by
\begin{equation}
  \label{eqn:approx-s-s-SDRR}
  K_{\scriptscriptstyle \textnormal{off}}\frac{\mathrm{d}\hat{\mathcal{P}}_{\scriptscriptstyle \infty}}{\mathrm{d}\lambda} = \left(\rho_{\scriptscriptstyle \infty}+J_{\scriptscriptstyle \textnormal{off}}\right) \hat{\mathcal{P}}_{\scriptscriptstyle \infty}(\lambda) - \frac{1}{2}\,\hat{\mathcal{P}}_{\scriptscriptstyle \infty}^{\hspace{0.5pt}2}(\lambda)-\frac{J_{\scriptscriptstyle \textnormal{on}}}{1 + \lambda},
\end{equation} which can be reduced to a special case of the associated Laguerre differential equation \cite{Abramowitz1965}, and its solution can be written in terms of the confluent hypergeometric function of the second kind $U$ \cite{Note3}, namely
\begin{equation}
  \label{eqn:solP}
  \hat{\mathcal{P}}_{\scriptscriptstyle \infty}(\lambda) = \frac{\displaystyle
  J_{\scriptscriptstyle \textnormal{on}}\,U\!\left(1-\kappa;\,1;\,\left(1+\lambda\right)\mathcal{J}\big/\kappa\right)
  }{\displaystyle
  K_{\scriptscriptstyle \textnormal{off}}\,U\!\left(-\kappa;\,0;\,\left(1+\lambda\right)\mathcal{J}\big/\kappa\right)},
\end{equation} where we define that $\kappa = J_{\scriptscriptstyle \textnormal{on}}\left[\hspace{1pt}2\hspace{1pt}K_{\scriptscriptstyle \textnormal{off}}\left(\rho_{\scriptscriptstyle \infty}+J_{\scriptscriptstyle \textnormal{off}}\right)\hspace{1pt}\right]^{-1}$ and $\mathcal{J} = \frac{1}{2}\,J_{\scriptscriptstyle \textnormal{on}}/K^{\hspace{0.5pt}2}_{\scriptscriptstyle \textnormal{off}}$. Since the solution in Eq.~(\ref{eqn:solP}) depends on the undetermined constant $\rho_{\scriptscriptstyle \infty}$, its value can be found by requiring the boundary condition $\rho_{\scriptscriptstyle \infty} = \hat{\mathcal{P}}_{\scriptscriptstyle \infty}(\lambda=0)$, which leads to a characteristic equation that needs to be numerically solved \cite{Note3}. Fig.~\ref{fig:3}\,(b) shows the numerical inversion of the Laplace transform in Eq.~(\ref{eqn:solP}) through a multi-precision computing algorithm \cite{Abate2004}. The functional form of the distribution is found to be  similar to the size-independent case, c.f.\ Eq.~(\ref{eqn:power-law}). However, as the value of $K_{\scriptscriptstyle \textnormal{off}}$ increases, the exponential cutoff $a_{\scriptscriptstyle C}$ of the distribution is significantly decreased, diminishing the power-law regime over which the system is scale-free, as well as reducing the average domain size, see Fig.~\ref{fig:3}\,(b).

Fortunately, the moments of the steady-state distribution can be computed from the derivatives of Eq.~(\ref{eqn:solP}) around the value of $\lambda=0$~\cite{Note3}. Moreover, the expression in Eq.~(\ref{eqn:solP}) can be expanded to first order in $\lambda_\star$, which allows us to find the linearized forms of the steady-state moments~\cite{Note3}. In particular, the total number-per-area of membrane domains and their area coverage can be written as follows: $\rho_{\scriptscriptstyle \infty}\approx\mathcal{Q}_{\scriptscriptstyle \infty}-J_{\scriptscriptstyle \textnormal{off}}-J_{\scriptscriptstyle \textnormal{on}}\hspace{1pt}K_{\scriptscriptstyle \textnormal{off}}/\left(J_{\scriptscriptstyle \textnormal{off}}\hspace{1.5pt}\mathcal{Q}_{\scriptscriptstyle \infty}\right)$ and $\phi_{\scriptscriptstyle \infty} \approx J_{\scriptscriptstyle \textnormal{on}}/J_{\scriptscriptstyle \textnormal{off}} - J_{\scriptscriptstyle \textnormal{on}}\hspace{1pt}K_{\scriptscriptstyle \textnormal{off}}\hspace{1pt}\!\left(J_{\text{on}} + 2\hspace{0.5pt}J_{\text{off}}^{\hspace{0.5pt}2}\right)\!/J^{\hspace{0.5pt}4}_{\scriptscriptstyle \textnormal{off}}$, respectively. 
Higher order moments can also be calculated, but their expressions become increasingly cumbersome \cite{Note3}, e.g. the second moment $\sigma_{\scriptscriptstyle \infty} \approx J_{\text{on}}\hspace{0.5pt}\!\left(J_{\text{on}} + 2\hspace{0.5pt}J_{\text{off}}^{\hspace{0.5pt}2}\right)\!/J_{\scriptscriptstyle \textnormal{off}}^{\hspace{0.5pt}3} - J_{\scriptscriptstyle \textnormal{on}}\hspace{1pt}K_{\scriptscriptstyle \textnormal{off}}\hspace{1pt}\!\left(6\hspace{0.5pt}J_{\scriptscriptstyle \textnormal{off}}^{\hspace{0.5pt}4} + 10\hspace{1pt}J_{\scriptscriptstyle \textnormal{on}}\hspace{0.5pt}J_{\scriptscriptstyle \textnormal{off}}^{\hspace{0.5pt}2} + 5\hspace{0.5pt}J_{\scriptscriptstyle \textnormal{on}}^{\hspace{0.5pt}2}\right)\!/J^{\hspace{0.5pt}6}_{\scriptscriptstyle \textnormal{off}}$. These statistical moments are in principle experimentally measurable, and thus it allows us to estimate the values of $J_{\scriptscriptstyle \textnormal{on}}$, $J_{\scriptscriptstyle \textnormal{off}}$, and $K_{\scriptscriptstyle \textnormal{off}}$ by simultaneously solving the above equations. This inference problem is illustrated in Fig.~\ref{fig:4}, where $J_{\scriptscriptstyle \textnormal{on}}$, $J_{\scriptscriptstyle \textnormal{off}}$, and $K_{\scriptscriptstyle \textnormal{off}}$ are selected such that they retrieve physiological values for the area-fraction and the mean domain size. Here, $K_{\scriptscriptstyle \textnormal{off}}$ is chosen to be considerably smaller than the other rates, so that the linear approximation still holds. Fig.~\ref{fig:4} shows how a fractional change in one of the moments affects the inferred recycling rates. 

In summary, we developed an out-of-equilibrium model for the in-plane membrane dynamics of domain structures, where their stability and characteristic sizes are mediated by the exchange of membrane components with the exterior. The dynamics and the steady-state values of the domain size distribution are studied within an aggregation model, subject to a size-dependent recycling scheme. Our analysis suggests a number of possible methods to experimentally test this model.

\begin{acknowledgments}
We acknowledge the stimulating discussions with Dr. P. Sens (Paris) and Dr. M. Rao (Bangalore), and funding from EPSRC under Grant No. EP/I005439/1 (M.S.T.).
\end{acknowledgments}

%

\end{document}